\documentclass[pre,aps,groupaddress,showpacs,twocolumn]{revtex4}
\usepackage{epsfig,amsmath,graphicx,amssymb,overpic}

\usepackage{graphicx}
\usepackage{dcolumn}
\usepackage{bm}
\usepackage{graphicx}
\usepackage{subfigure}
\def\be{\begin{equation}}
\def\ee{\end{equation}}
\def\bee{\begin{eqnarray}}
\def\ene{\end{eqnarray}}
\def\bes{\begin{subequations}}
\def\ees{\end{subequations}}

\newcommand{\br}{{\bf r}}
\newcommand{\bv}{{\bf v}}

\newcommand{\bc}{{\bf c}}
\newcommand{\ba}{{\bf a}}

\newcommand{\bb}{{\bf b}}

\begin{document}
\title{Exact solutions to three-dimensional generalized
 nonlinear Schr\"odinger equations with varying potential and nonlinearities}
\author{Zhenya Yan$^{1,2}$}
\email{zyyan@mmrc.iss.ac.cn}

\author{V. V. Konotop$^{1,3}$}
\email{konotop@cii.fc.ul.pt}

\affiliation{$^1$Centro de F\'{\i}sica Te\'orica e Computacional,
Universidade de Lisboa, Complexo Interdisciplinar, Avenida
Professor Gama Pinto 2, Lisboa 1649-003, Portugal \\
$^2$Key Laboratory of Mathematics Mechanization, Institute of
Systems Science, AMSS, Chinese Academy of Sciences, Beijing
100080, China \\
$^3$Departamento de F\'{\i}sica, Universidade de Lisboa, Campo
Grande, Ed. C8, Piso 6, Lisboa 1749-016, Portugal \vspace{0.1in}}

\date{\vspace{0.1in} 4 May 2009, Phys. Rev. E  {\bf 80}, 036607 (2009)}

\begin{abstract}

It is shown that using the similarity transformations, a set of three-dimensional p-q nonlinear Schr\"odinger (NLS)
 equations with inhomogeneous coefficients can be reduced to one-dimensional stationary NLS equation
 with constant or varying coefficients, thus allowing for obtaining exact localized and periodic wave solutions.
 In the suggested
  reduction the original coordinates in the (1+3)-space are mapped into a set of one-parametric coordinate surfaces,
  whose parameter plays the role of the coordinate of the one-dimensional equation. We describe the algorithm
   of finding solutions and concentrate on power (linear and nonlinear) potentials  presenting a number of case
   examples. Generalizations of the method are also discussed.

\end{abstract}

\maketitle


\section{Introduction}

The nonlinear Schr\"odinger (NLS) equation is a key model describing wave processes in weakly dispersive and weakly
nonlinear media~\cite{Sulem}. It has numerous physical applications and this universality stimulates a great deal
of attention  devoted to search of exact solutions of the generalized NLS  models which include coefficients
depending on  spatial and temporal variables, i.e.  describing wave dynamics in inhomogeneous media. The first
 exact results were obtained for the one-dimensional (1D) NLS equation using the inverse scattering
 technique~\cite{BLR} and later on generalized for the respective discrete models~\cite{Discrete} and to NLS with
 random coefficients~\cite{Bisieris}.
 Applications of the NLS equation in fiber optics have stimulated further studies of the integrable
 inhomogeneous models leading to the concepts of self-similar solitons and
 non-autonomous solitons put forward in Refs.~\cite{Serkin}.

Very recently the interest in exact solutions of the NLS equations
with inhomogeneous coefficients was stimulated by its applications
in the mean-field theory of Bose-Einstein condensate (BEC), where
it is also known as Gross-Pitaevskii equation~\cite{PitStrin}.
Except a few special cases~\cite{general_inhom}, the inhomogeneous
NLS equation in the BEC applications appears to be nonintegrable,
either due to due its two-- or three--dimensional  nature or  in
quasi-1D approximation due to the respective inhomogeneous terms
(such as parabolic potentials and nonlinear inhomogeneities).
Therefore new approaches, based on the self-similar
transformations, have been developed. In particular, exact
solutions in 1D NLS equations with stationary inhomogeneous
coefficients were constructed in Refs.~\cite{nontri,BK,Juan1},
solutions of the NLS model with coefficients depending on time and
space variables were
 considered in~\cite{Juan2,space-time}. A $d$-dimensional  NLS equation with varying coefficients was considered in~\cite{PTK},
 where the lens transformations allowed for its reduction  to the $d$-dimensional model with constant coefficients,
 whose dynamical properties are known. Nontrivial solutions of the cubic-quintic NLS equation with a
 periodic potential were also considered in~\cite{5nls}. The solutions of the cubic-quintic NLS model with
 coefficients depending on time and space variables were addressed in~\cite{Juan3}.

The present paper aims to report a possibility of generating exact
solutions of a generalized three-dimensional (3D) NLS equation
with inhomogeneous coefficients employing the self-similar
reduction. Unlike in the previous studies we implement {\em
reduction of the spatial dimension of the system}. More
specifically we consider the
 mapping of the  coordinate surfaces in the 3D space into a one-parametric family, where the quantity parameterizing
 the surface family serves as a variable of the 1D NLS equation with constant coefficients, thus  allowing for
 immediate indication of a great number of exact solutions.

The solutions we will obtain have nontrivial phases, thus
representing the hydrodynamic flows in specially created
potentials. Bearing this in mind and taking into account that the
explored potentials (parabolic and quartic)  are typical for the
BEC applications, we refer to such solutions also as to (exact)
{\em flows} and employ the terminology widely accepted in the BEC
theory.

The paper is organized as follows. In Sec. II, we describe the
similarity transformation. In Sec. III, we
focus on the amplitude and phase surfaces associated to the introduced transformations and present  stationary solutions.  Sec. IV is devoted to
time-dependent cases. In Sec. V, we study the generalizations of the theory, including reduction to the 1D equations
 with inhomogeneous coefficients but allowing for exact solutions. The outcomes are summarized in the
 Conclusion.

\section{Similarity reductions and solutions}

\subsection{The similarity reduction}

We concentrate on the 3D inhomogeneous p-q NLS equation with
varying coefficients
\begin{eqnarray}
\label{5NLS}
 i\frac{\partial \psi}{\partial t}=-\frac{1}{2}\nabla^2\psi &\!\!\!+&\!\!\! v(\br,t)\psi
 \nonumber \\
 &\!\!\!+&\!\!\!
 \left[g_{p}(\br,t)|\psi|^{p-1}+g_{q}(\br,t)|\psi|^{q-1}\right]\psi,
 \quad
 \end{eqnarray}
where $\psi\equiv\psi(\br, t)$,\ $\br\in\mathbb{R}^3$, $\nabla
\equiv(\partial_x,\partial_y,\partial_z)$, $q>p\geq 3$  are
integers,  the linear potential $v(\br,t)$  and the nonlinear
coefficients $g_{p,q}(\br,t)$ are all real-valued functions of
time and spatial coordinates. This model contains many special
types of nonlinear equations with varying coefficients such as the
cubic NLS equation,  the cubic-quintic NLS model, the generalized
NLS model, etc.

Following the procedure suggested in \cite{Juan2, Yanps07} we
search for a transformation connecting solutions of
Eq.~(\ref{5NLS}) with those of the stationary p-q NLS equation
with constant coefficients \bee
\mu\Phi=-\Phi_{\eta\eta}+G_{p}|\Phi|^{p-1}\Phi+G_{q}|\Phi|^{q-1}\Phi.
\label{ODE} \ene Here $\Phi\equiv \Phi(\eta)$ is a function of the
only variable $\eta\equiv \eta(\br,t)$ whose relation to the
original variables   $(\br,t)$ is to be determined, $\mu$ is the
eigenvalue of the nonlinear equation, and $G_{p,q}$ are constants.
Since both $|G_p|$ and $|G_q|$ can be scaled out (by the proper
renormalization of the amplitude, of $\mu$, and of the
``coordinate" $\eta(\br, t)$) without loss of generality the
consideration will be restricted to the cases where $G_p=0,\pm 1$
and $G_q=0, \pm 1$.

In order to control the boundary conditions at the infinity we
impose  the  natural constraints
\bee \label{constrain1} \eta\to
0\quad\mbox{at}\quad r\to0\quad \mbox{and}\quad
\eta\to\infty\quad\mbox{at}\quad r\to\infty. \ene

Thus we consider the general similarity transformation \bee
\psi(\br,t)=\rho(\br, t)e^{i\varphi(\br,t)}\Phi(\eta(\br,t)),
\label{Tran} \ene where $\varphi(\br,t)$ is a real-valued function
and $\rho(\br, t)$ is a nonnegative function of the indicated
variables, the both are to be determined. Requiring $\Phi(\eta)$
to be real (without loss of generality) and to satisfy Eq.
(\ref{ODE}), we substitute the ansatz (\ref{Tran}) into
Eq.~(\ref{5NLS}) and after simple algebra obtain the set of
equations
\begin{subequations}
\label{sys} \bee \label{sys1}
 && \nabla\cdot(\rho^2\nabla\eta)=0, \\
 \label{sys2}
 && (\rho^2)_t+\nabla\cdot(\rho^2\nabla\varphi)=0,\\
 \label{sys3}
 &&\eta_t+\nabla\varphi\cdot\nabla\eta=0, \\
 \label{sys4}
 &&2g_j(\br,t)\rho^{j-1}-G_j |\nabla\eta|^2=0 \ \ (j=p,\,q),  \\
 \label{sys5}
 &&2v(\br,t)+\mu
 |\nabla\eta|^2+|\nabla\varphi|^2-\rho^{-1}\nabla^2\rho+2\varphi_t=0.
 \qquad
\ene
\end{subequations}
These  equations lead to several immediate conclusions.
   First, it follows
from (\ref{sys4}) that $g_{p,q}(\br,t)$ are sign definite, and
$G_j={\rm sign}\{g_{j}(\br,t)\}$. Moreover, comparing the
 equations in (\ref{sys4}) for $j=p$ and $j=q$ we find that
either $|g_{p} |=\rho^{q-p}|g_{q}|$, or one of
the nonlinear coefficients is zero, i.e. either
$|g_{p}|\equiv 0$ or $|g_{q}|\equiv 0$.
Respectively, we define the function $g(\br,t)\equiv
2g_{j} \rho^{j-1}/G_{j}$,
 where $j=p,q$.

To solve the system (\ref{sys}) explicitly,  we first consider the
special case of $\rho(\br,t) $ depending only on time $t$, i.e.
$\rho(\br,t)\equiv\rho(t)$.
 Then the  system (\ref{sys}) is simplified:
\begin{subequations}
\label{set} \bee
 \label{set1}
 && \nabla^2\eta=0,
\\  \label{set2}  &&
 2\rho_t+\rho\nabla^2\varphi=0,
 \\  \label{set3}  &&
 \eta_t+\nabla\varphi\cdot\nabla\eta=0, \\
 \label{set4}
 && g(\br,t)- |\nabla\eta|^2=0, \\
 \label{set5}
 &&2v(\br,t)+\mu |\nabla\eta|^2+|\nabla\varphi|^2+2\varphi_t=0.
\ene
\end{subequations}

As it is clear, the equations in the system (\ref{set}) are not
compatible with each other in the case of arbitrary linear and
nonlinear potentials.  One however can pose the problem to find
  functions $v(\br,t)$ and $g(\br,t)$, for which the mentioned
system becomes solvable.  This leads us to the procedure which can
be outlined as follows.

\begin{itemize}
    \item First, one solves Eqs. (\ref{set1})-(\ref{set3}) [or Eqs. (\ref{sys1})-(\ref{sys3})]  subject to the boundary
    conditions (\ref{constrain1})
    what gives the functions $\eta(\br,t)$ and $\varphi(\br,t)$.
    \item Second, one considers Eqs. (\ref{set4}) and (\ref{set5}) [or Eqs. (\ref{sys4}) and (\ref{sys5})] as
     definitions
     for the functions $v(\br,t)$ and
    $g(\br,t)$ through the already known $\eta(\br,t)$ and $\varphi(\br,t)$.
    \item Third, using one of the known solutions of the stationary p-q NLS equation (\ref{ODE}) and
    the similarity transformation (\ref{Tran}) one can construct
    the analytical solutions of Eq. (\ref{5NLS}).
\end{itemize}

The last step is trivially performed, taking into account that $\Phi$ is real, giving a solution in the implicit form
\begin{eqnarray}
\label{implicit}
    \eta=\int
    d\Phi\left[C-\mu\Phi^2+\frac{2G_p}{p+1}\Phi^{p+1}+\frac{2G_q}{q+1}\Phi^{q+1}\right]^{-\frac{1}{2}},
\end{eqnarray}
where $C$ is an integration constant.

In the next
sections we implement the describe approach for a number of
particular cases, relevant to the physical applications. Before that however we briefly address the issue of the integrals of motion.

\subsection{One integral of motion}

Integrals of motions generally appear to be the most important
characteristics (either physical or mathematical) of the motion.
The simplest conserved quantity for an $L^2$ integrable solution
of the NLS equation -- the number of particles, for the whole
space is not defined in the case at hand, since the solutions we
are dealing with do not decay at the infinity. Instead, however,
as it is customary for the classical hydrodynamics  dealing with
flows we define a number of particles in a simply connected
bounded volume $U\in \mathbb{R}^3$ which consists of the same
"particles" and thus moves  with the "fluid" described by the NLS
equation (\ref{5NLS}), i.e.  $U\equiv U(t)$:
\begin{eqnarray}
\label{N}
N_U=\int_{U(t)}|\psi(\br,t)|^2 d\br.
\end{eqnarray}
Under the term ``particles" we understand an infinitesimal volume
of the fluid moving with the velocity $\bv\equiv\nabla \varphi$.
Then, the solutions considered in the present paper possess the
properties of the conservation of $N_U$:
\begin{eqnarray}
\label{N_cons}
\frac{dN_U}{dt}=0.
\end{eqnarray}
This formula represents nothing else than the well known transport
formula of the conventional hydrodynamics~\cite{hydro}.

In order to prove (\ref{N_cons}) we first observe
 that it follows from Eqs. (\ref{5NLS}) and
(\ref{Tran}) that
 \bee
   \frac{\partial}{\partial t}|\psi|^2=-\nabla[(\rho\Phi)^2\nabla \varphi]\,.
 \ene
This equation combined with the transport formula, results in the set of equalities
 \bee
  \displaystyle\frac{d}{dt} \int_{U(t)} |\psi|^2d\br
  =\displaystyle \int_{U(t)}\left[\frac{\partial}{\partial t}|\psi|^2+\nabla(|\psi|^2\cdot {\bf v})\right]d\br
    \nonumber \\
    = \displaystyle\int_{U(t)}\left[\frac{\partial}{\partial t}|\psi|^2+\nabla[(\rho\Phi)^2\nabla
  \varphi]\right]d\br  = 0.
 \ene

In spite of the apparent complexity of the solutions (\ref{Tran}), the obtained conservation of the number of particles in a material volume moving with the fluid, does not appear too surprising. Indeed, one can take into account that the symmetries of the system used to construct the solutions are based on reduction to the {\em stationary} model (\ref{ODE}).

\section{Surfaces and stationary solutions}
\label{sec:stationary}

We start with the stationary solutions of Eq. (\ref{5NLS}),
$\rho_t=\eta_t=\varphi_t=0$,  imposing even more strong constrain
on $\rho$ requiring it to be $\br$-independent constant. Then without loss of
generality we set $\rho=1$. Also now the linear and nonlinear
potentials do not depend on time $t$, i.e. $v(\br,t)\equiv v(\br)$
and $g(\br,t)\equiv g(\br)$  [recall that now it is mandatory to
have $g(\br) >0$].

Introducing the notation $u(\br)\equiv -2v(\br)-\mu g(\br)$ we
rewrite the system (\ref{set}) in the stationary case as
\begin{subequations}
\label{set_stat} \bee \nabla^2\eta=0,\quad \nabla^2\varphi=0,\quad
\nabla\eta\cdot\nabla\varphi=0, \label{set_stat_1}
\\
\label{set_stat_2} |\nabla\eta|^2=g(\br),\quad
|\nabla\varphi|^2=u(\br). \quad\quad\quad \ene
\end{subequations}
 It follows from the second of
Eqs. (\ref{set_stat_2}) that $u(\br)\geq 0$, and hence one must
require $v(\br)\leq -\frac{1}{2}\mu g(\br)$.

\subsection{Amplitude and phase surfaces. The potentials.}

Now we consider surfaces of the constant amplitude and phase, i.e.
\bee \label{eta} \eta(\br)=\eta_0=\mbox{const}
\quad\mbox{and}\quad \varphi(\br)=\varphi_0=\mbox{const}. \ene
First, we observe that the only singular points of the amplitude
surfaces occur where the system becomes linear, i.e. where
$g(\br)=0$ since otherwise $\nabla\eta(\br)\neq 0$.  Next, having
defined one of the surfaces (we will always start with the
coordinate surface), the last equation in Eq.(\ref{set_stat_1})
appears to be an important constraint of the definition of the
other surface (in our case it will be the phase surface).

In fact, the first two equations in Eq. (\ref{set_stat_1}) imply
that the amplitude $\eta(\br)$ and phase $\varphi(\br)$ belong to
the kernel $L=\{f(\br)| \nabla^2f(\br)=0\}$ of the Laplace
operator $\nabla^2$, which are the harmonic functions and form a
function space in $\mathbb{R}$. It follows from the last equation
inn Eq. (\ref{set_stat_1}) that the dot product of the gradients
of amplitude $\eta(\br)$ and phase $\varphi(\br)$ is to zero. That
is to say,  $\eta(\br)$ and $\varphi(\br)$ are harmonic functions
and their gradients are orthogonal.

In what follows we restrict our consideration to the finite power
($N$-order) surfaces, i.e. depending on terms like
$x^{n_1}y^{n_2}z^{n_3}$ with $n_j$ being finite positive integers
such that $\max\{n_1+n_2+n_3\}= N<\infty$. This allows us to list
in the Table~\ref{surface} all admissible coordinate  and   phase
surfaces  which appear to be not higher than the third-order, i.e.
$N\leq3$ (the second and third columns, respectively). The number
of surfaces is limited by the irreducible cases, i.e. to the
surfaces that cannot be transformed to each other by proper change
of the coordinates leaving the Laplacian invariant. In all the
cases the stationary phase is set zero. Notice that if the phase
$\varphi(\br)$ is chosen as a constant, then
Eq.~(\ref{set_stat_1}) is reduced to the single Laplace equation
$\nabla^2\eta(\br)=0$ for  the amplitude $\eta(\br)$ whose
solutions are just the harmonic functions (in what follows we will
not consider them).


\begin{table*}[ht]
\caption{Admissible coordinate and phase surfaces and the respective linear
and nonlinear potentials. To reduce the number of constants, those that can be scaled out
by change of the coordinate units are set to one. All the constants left are real.}
 \begin{tabular}{lllll} \hline\hline \\ [-2.0ex]
 Case & Amplitude surface \quad\quad\quad\quad\quad & Phase surface \quad\quad\quad\quad\quad\quad\quad
 & Linear potential $v(\br)$ \quad\quad\quad &  Nonlinear potential  $g(\br)$
 \\ [1.0ex] \hline \\ [-2.0ex]
 I & $\eta(\br)=\bc\cdot\br$  &  $\varphi(\br)=\ba\cdot\br $ & $-(\mu|\bc|^2+|\ba|^2)/2 $
          &  $|\bc|^2$  \\
   & (plane, $|\bc|\neq 0$)  & (plane, $|\ba|\not=0,\ \bc\cdot\ba=0$) & (constant) & (constant) \\ [1.0ex]
     \hline \\ [-2.0ex]
 II & $\eta(\br)=x +c (y^2-z^2) $ & $\varphi(\br)=2ayz$   &   $-2(\mu c^2+a^2)(y^2+z^2)-\mu/2$ \
     & $4c^2(y^2+z^2)+1$   \\
   & (hyperbolic paraboloid) \quad  & (hyperbolic cylinder)
   &  (elliptic cylinder) & (elliptic cylinder) \\  [1.0ex] \hline \\ [-2.0ex]
 III & $\eta(\br)= cx^2+(1-c)y^2-z^2 $ \quad\quad & $\varphi(\br)=2axyz$    & $-2\mu [c^2x^2+(1-c)^2y^2+ z^2]$ &
     $4[c^2 x^2+(1-c)^2y^2+ z^2]$  \\
    & (hyperboloid of one ($c\in (0, 1)$)  \qquad    & (third order surface) & \quad $-2a^2(y^2z^2+x^2z^2+x^2y^2)$
    &  (real ellipsoid) \\
    & \ or two ($c>1$) sheets and & & \ (fourth order surface) &   \\
    & \ hyperbolic cylinder ($c=1$)) &&&
    \\ [1.0ex] \hline \\ [-2.0ex]
 IV & $\eta(\br)=xyz $  & $\varphi(\br)=a_1x^2+a_2y^2-a_3z^2$   & $-\mu(y^2z^2+x^2z^2+x^2y^2)/2$ \quad &
    $ y^2z^2+x^2z^2+x^2y^2$  \\
     & (third order surface)     & (similar to $\eta(\br)$ in the     & \quad $-2 (a_1^2x^2+a_2^2y^2+a_3^2z^2)$ &
       (fourth order surface) \\
    & & Case III, $a_3=a_1+a_2$) &  (fourth order surface) & \\
     [1.0ex]
  \hline\hline
\end{tabular}
\label{surface}
\end{table*}

  All solutions shown in Table~\ref{surface} describe irrotational flows: $\nabla\times \bv(\br)=0$ where as above
  $\bv(\br)=\nabla\varphi(\br)$.
  Another feature to be emphasized is that  the coordinate and phase surfaces, are defined by the linear part of
  the evolution equation
  (i.e. by the linear dispersion relation of the medium). They generate linear and nonlinear potentials.
  In practice however this picture is inverted: for a given linear and nonlinear potentials (created, say, experimentally)
  one has to introduce an appropriate
  coordinate surface. In this last sense the two last column in the Table~\ref{surface} indicate the types of
  the linear and nonlinear potentials for which the mapping $(x,y,z)\to\eta$ is possible.
  Namely we observe that   all the obtained solutions  require specific quadratic and quartic linear and nonlinear
  potentials ($v(\br),\ g(\br)$).

 The Case I in Table~\ref{surface}  describes a flow with the constant velocity $\bv(\br)=(a_1,a_2,a_3)$ which is
 generated
  by the constant nonlinearity and linear potentials (the latter, obviously, can be removed). This is the trivial
  case of line-solutions, which by simple rotation of the coordinates are reduced to solutions depending only on one
  coordinate (say, $x$) and independent on other coordinates. In what follows will not consider them.

\begin{figure}
 \begin{center}
  {\scalebox{0.5}[0.5]{\includegraphics{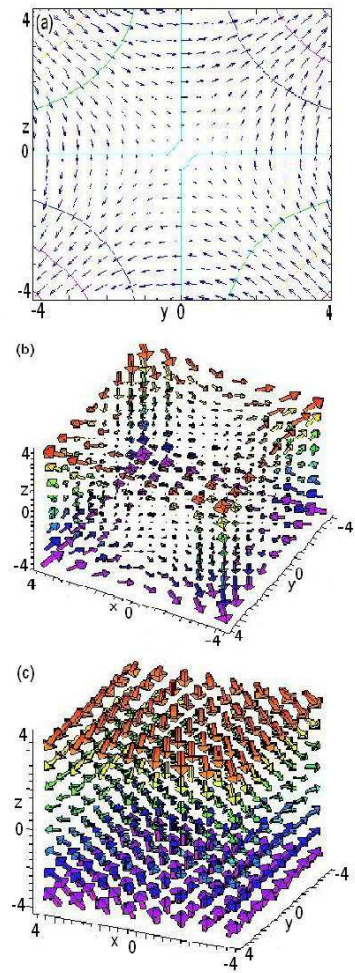}}} \qquad\qquad\qquad\qquad\qquad
   \hspace{1in}
   \end{center}
 \vspace{-0.4in}\caption{ (Color online) The velocity fields $\bv =\nabla\varphi $ corresponding to the
 phases
listed  in Table~\ref{surface} for  $a=a_{1,2}=1$. (a)
($y, z$)-plane for Case II, (b) 3D-space for Case III ,
(c) 3D-space for Case IV.} \label{fig:surf23}
\end{figure}

 The Case II in Table~\ref{surface} describes flows which are outgoing  in the first and third quadrants in the $(y,z)$-plane,  and  incoming in the second and forth quadrants, with respect to the $x$-axis (see Fig.~\ref{fig:surf23}a). The velocity is given by $\bv(\br)=(0,2az,2ay)$. Such flows
 is generated by the parabolic linear (either repulsive or attractive) and parabolic nonlinear  potentials. Being symmetric they preserve
 the total number of particles.

 The Case III in Table~\ref{surface} is a 3D flow with the velocity $\bv(\br)=(2ayz, 2axz, 2axy)$, which is
 generated by the quartic linear and quadratic nonlinear potentials (see Fig.~\ref{fig:surf23}b).
 In  each coordinate plane, i.e. $(x, y)-$, $(x, z)-$, and $(y, z)-$plane, they
 all describe outgoing  in the first and third quadrants,  and
 incoming  in the second and forth quadrants, flows.

 The Case IV in Table~\ref{surface} is a 3D flow with the velocity $\bv(\br)=(2a_1x, 2a_2y, -2(a_1+a_2)z)$,
  which are
 generated by the quartic linear and quartic nonlinear potentials (see Fig.~\ref{fig:surf23}c).

In all the cases the types of the nonlinear interactions are determined by the constants $G_{p,q}$: they are
attractive at $G_{p,q}>0$ and repulsive at $G_{p,q}<0$. Meantime the linear potential, which depends on the
chemical potential is related to the type of the solution (as the different types of the solutions exist for
different signs of $\mu$, see below). For $\mu<0$ the linear potentials can change the sign of their curvatures in different points
of the space.

\subsection{Solutions: cubic NLS equation}

Following the algorithm described above, in order to construct the
exact solutions of Eq. (\ref{5NLS}), as the last step we have to
address solutions of  Eq.~(\ref{ODE}) [i.e. the formula
(\ref{implicit})]. They depend on the particular choice of the
model. In the present work we consider two the most relevant
physical cases. First we concentrate  on  the standard cubic NLS
equation and after that we make comments on the cubic-quintic model. Thus starting with the case $p=3$, $g_q(\br)\equiv 0$ and hence  $G_q=0$ we have to deal with the NLS equation
$\mu\Phi=-\Phi_{\eta\eta}+G_3\Phi^3$.
 The respective periodic and localized solutions are very well known. Below we consider the simplest
 ones for attractive and repulsive nonlinearities   $G_3$.

\subsubsection{Attractive nonlinearity $G_3=-1$}

\begin{figure}
 \begin{center}
\hspace{-0.15in}
  {\scalebox{0.442}[0.5]{\includegraphics{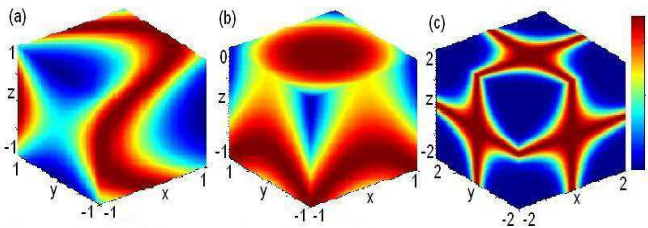}}}
  \end{center}
 \vspace{-0.2in}\caption{(Color online)
 Cross-sections of the density distribution of the ``bright soliton"
 $|\psi_{bs}(\br)|^2$ with $\eta$ given in Table \ref{surface} for  $\mu=-1$.
 (a) $c=1$ with $\eta$ for Case II, (b) $c =0.5$ with $\eta$
for Case III; here the cross-section at
$z=0$ shows the peak intensity,  (c) $\eta$ for Case IV.} \label{fig:f3b}
\end{figure}

Now the simplest stationary nontrivial solution is the  NLS bright
soliton, which gives  $\psi_{bs}(\br)=\sqrt{-2\mu}\ {\rm
sech}[\sqrt{-\mu}\,\eta(\br)]\exp[i\varphi]$, where $\mu<0$
and the amplitude $\eta(\br)$ and phase $\varphi(\br)$ are defined
by Table \ref{surface}.
 In Fig.~\ref{fig:f3b}, we display the cross-sections of the intensity $|\psi_{bs}(\eta(\br))|^2$
 of the bright soliton solution for different types of amplitude surfaces $\eta(\br)$ given in Table~\ref{surface}. We emphasize that while we use the   solitonic terminology referring to the bright solitons, the respective solutions are not decaying in the 3D case we are interested in as this is illustrated by Fig.~\ref{fig:f3b} (this comment on the usage of the 1D terminology is also relevant to all other solutions considered below).

We also observe that the described solutions allow for direct
generalization to the p-NLS case of arbitrary even $p\geq 4$ and
$G_{q}=0$ for which Eq. (\ref{ODE}) becomes $
\mu\Phi=-\Phi_{\eta\eta}+G_{p}\Phi^{p}$. The bright soliton
solution of Eq. (\ref{5NLS}) obtained from (\ref{implicit})
corresponds to $G_{p}=-1$ and $\mu=-4\omega^2(p-1)^{-2}$ and is  given
by $ \psi_{pbs}(\br)=\left[\frac{\omega\sqrt{2(p+1)}}{p-1}\, {\rm
sech}(\omega\eta(\br))\right]^{2/(p-1)}e^{i\varphi(\br)}$, where
$\omega$ is a constant.

\begin{figure}
 \begin{center}
\hspace{-0.18in}
  {\scalebox{0.447}[0.5]{\includegraphics{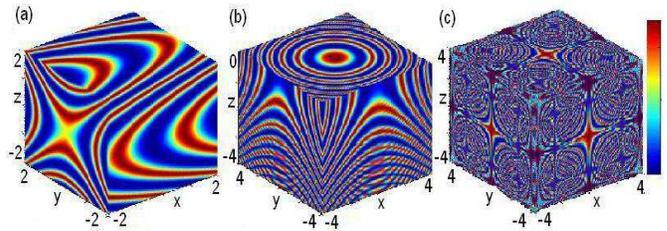}}}
  \end{center}
\vspace{-0.2in}\caption{(Color online) Cross-sections of the
density distribution of the   cn-wave solution
$|\psi_{cn}(\br)|^2$ with $\eta$ given in Table \ref{surface} for
 $\mu=-1$. (a) $c=1$ with $\eta$ for Case II,
(b) $c =0.5$ with $\eta$ for Case III,   peak intensity at the center is shown by
the cross-section at $z=0$,  (c) $\eta$ for Case IV. In all the
panels $k=0.8$.} \label{fig:f3p}
\end{figure}

Next we address the solutions of the 3D model (\ref{5NLS}) generated by the periodic cn-wave
 solution of the cubic NLS equation
 \bee
  \label{3p}
   \psi_{cn}(\br)=\sqrt{\frac{2\mu k^2}{1-2k^2}}\, {\rm cn}\left[\sqrt{\frac{2\mu}{1-2k^2}}\eta(\br) ,
   k\right]
    e^{i\varphi(\br)},
 \ene
 where $k\in [0, \ 1]$ is the modulus of the Jacobi elliptic function and   $\mu$ satisfies
 the condition
$\mu(1-2k^2)>0$, i.e. $\mu<0, \ 1/\sqrt{2}<k<1$ or $\mu>0,\
0<k<1/\sqrt{2}$. As before the amplitude surface $\eta(\br)$ and phase
surface $\varphi(\br)$ are defined by Table \ref{surface}. Examples of the mentioned solutions are shown in
Fig.~\ref{fig:f3p}.

There are two features of the solutions to be emphasized here. First, being periodic in $\eta$ the solutions
 are not periodic in the real 3D space. However as the ``trace" of the periodicity, in Fig.~\ref{fig:f3p} one
  observes repeated domains of maxima and minima of the density, unlike in  Fig.~\ref{fig:f3b}, where in each
  cross-section one observes only one curve corresponding to the maximum of the density (which follows the projection of the amplitude surface on the plane  of the chosen cross-section).

\subsubsection{Repulsive nonlinearity $G_3=1$}

\begin{figure}
 \begin{center}
\hspace{-0.18in}
  {\scalebox{0.447}[0.5]{\includegraphics{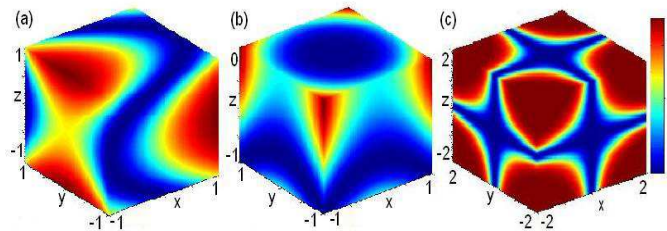}}}
  \end{center}
\vspace{-0.2in}\caption{(Color online) Cross-sections of the
density distribution of the ``dark soliton" $|\psi_{ds}(\br)|^2$
with $\eta$ listed in Table \ref{surface} for $\mu=2$. (a) $c=1$
with $\eta$ for Case II, (b) $c =0.5$ with $\eta$ for Case III, the minimum intensity  is shown in the cross-section with $z=0$, (c)
$\eta$ for Case IV. } \label{fig:f3d}
\end{figure}

This is the case where $\mu$ is positive.
Now one has the dark soliton solution of Eq.~(\ref{5NLS}):
$\psi_{ds}(\br)
=\sqrt{\mu}\tanh[\sqrt{\mu/2}\,\eta(\br)]\exp(i\varphi(\br))$.
 The respective intensity profiles $|\psi_{ds}(\br)|^2$ are represented in Fig.~\ref{fig:f3d} for different
 types of amplitude and phase surfaces from Table~\ref{surface}.

Now the linear potential is repulsive in the whole space and in center of coordinates the density becomes zero (see Fig.~\ref{fig:f3d}b.

\begin{figure}
 \begin{center}
\hspace{-0.18in}
  {\scalebox{0.447}[0.5]{\includegraphics{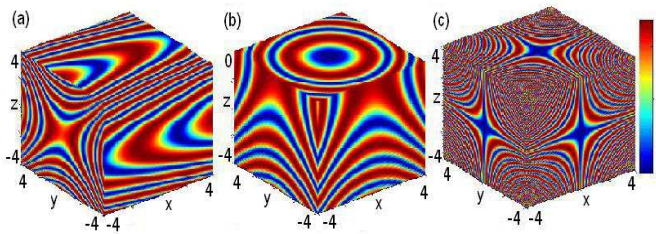}}}
  \end{center}
\vspace{-0.2in}\caption{(Color online) Cross-sections of the
density distribution of the sn-wave solution $|\psi_{sn}(\br)|^2$
with $\eta$ given in Table~\ref{surface} for
 $\mu=2$. (a) $c=1$ with $\eta$ for Case II,
(b) $c =0.5$ with $\eta$ for Case III, the minimum intensity is shown in the cross-section at $z=0$,  (c) $\eta$ for Case IV. In
all the panels $k=0.8$.} \label{fig:f3sn}
\end{figure}

 One can also construct ``periodic" sn-wave solutions
\bee
  \label{3pp}
   \psi_{sn}(\br)=\sqrt{\frac{2\mu k^2}{1+k^2}}\ {\rm sn}\left[\sqrt{\frac{\mu}{1+k^2}}\, \eta(\br) ,
   k\right]
    e^{i\varphi(\br)}
\ene
where we used the
 periodic sn-wave solution of the NLS equation with a positive chemical potential $\mu$. These solutions  are depicted in  Fig.~\ref{fig:f3sn},

For other possible types of the exact flows generated by the of cubic NLS equations see e.g.~\cite{Yanps}.

\subsection{Solutions: cubic-quintic NLS equation}

Now we briefly discuss the  cubic-quintic NLS equation, i.e. $p=3, \ q=5$
for which Eq.~(\ref{ODE}) becomes
$
\mu\Phi=-\Phi_{\eta\eta}+G_3\Phi^3+G_5\Phi^5.
$
Its solutions for the condition $G_3G_5<0$ are also known (some nontrivial examples
  are listed  in Table~\ref{CQsolution} given in Appendix~\ref{cub_quint};  for the methods of
construction of the solutions see also Refs.~\cite{Yanpla,Yanps}
for more details).  While the amplitude and the phase surfaces are
now the same as in the case of the cubic
 NLS equation, the density distribution is described by different periodic and localized functions. We in particular emphasize possibility of the algebraic solutions, like the ones given by the cases 4 and 7 in Table~\ref{CQsolution}.

\section{Time dependent amplitudes and phase surfaces, and potentials}

So far we have considered  stationary solutions. Now we allow $\rho(\br, t)$, $\eta(\br, t)$ and
$\varphi(\br, t)$ to depend on  spatial and temporal variables. As
before we focus on the finite power surfaces, i.e. depending on
terms like $ f_{n_1n_2n_3}(t)x^{n_1}y^{n_2}z^{n_3}$ with $n_j$
being finite positive integers such that $\max\{n_1+n_2+n_3\}=
N<\infty$ and $f_{n_1n_2n_3}(t)$ being functions on time $t$.
In what follows we consider all admissible coordinate  and phase
surfaces which appear to be not higher than the third-order, i.e.
$N\leq3$.

\subsection{Plane surface depending on time}

The first nontrivial result is obtained for moving plane surfaces
(which in the stationary case was reduced to the trivial
1D case).  To this end, based on the Case I of the
Table~\ref{surface} and  Eqs. (\ref{set1})-(\ref{set3}) we
consider  $\eta$ parameterizing moving plains
 \bee
   \eta(\br, t) =\bc(t)\cdot\br
 \label{varable1a}
 \ene
 where $\bc(t)=(c_x(t),c_y(t),c_z(t))$
is an arbitrary vector functions of time $t$ subject to the only
constrain $c_x(t)c_y(t)c_z(t)>0$ for any positive time $t>0$ (this constrain, as well as similar conditions below are imposed only for the sake of simplicity). The
nontrivial phase now reads
 \bee
   \varphi(\br,t)&=&\br\hat\Omega(t)\br +\ba(t)\cdot\br
\label{varable1b}
\ene
where we have introduced the diagonal time-dependent 3$\times$3 matrix $\hat\Omega=$diag$(\Omega_x,\Omega_y,\Omega_z)$ with
$\Omega_{\sigma}=-\dot{c}_{\sigma}(t)/(2c_{\sigma}(t))$ (hereafter
 $\sigma=x, y, z$) and $\ba (t)=(a_x(t), a_y(t), a_z(t))$
is a time dependent vector-function, such that the condition
 $ \bc(t)\cdot\ba(t)=0$ is satisfied.
   Now, from Eqs. (\ref{set4}) and (\ref{set5}) we  obtain $\rho(t)=\sqrt{c_x(t)c_y(t)c_z(t)},$
 \bee
 \nonumber
v(\br,t) =\br\hat A(t)\br
   +\bb (t)\cdot \br -\frac 12 (\mu |\bc(t)|^2+|\ba(t)|^2)  ,
\ene and $g(\br,t)=|\bc(t)|^2$. Here we have defined the diagonal
time-dependent 3$\times$3 matrix $\hat A=$diag$(A_x,A_y,A_z)$ with
the entries \bee \label{A}
 A_{\sigma}
 =\frac{\ddot{c}_{\sigma}(t)}{2c_{\sigma}(t)}-\frac{\dot{c}_{\sigma}^2(t)}{c_{\sigma}^2(t)},
 \ene
 and the vector function $\bb(t)=(b_x,b_y,b_z)$ with
 \bee
 \label{B}
b_{\sigma}=\frac{\dot{c}_{\sigma}(t)a_{\sigma}(t)}{c_{\sigma}(t)}-\dot{a}_{\sigma}(t).
\ene

The described solution can be realized using the time-dependent linear potential. The nontrivial
 effect arising in such a geometry is that the phase surface becomes of the second order and thus the solution at hand is characterized by the inhomogeneous in space and dependent on time
 velocity field.

\subsection{Paraboloid depending on time}

Next we consider the generalization of the parabolic Case II  from
the Table~\ref{surface} for which the amplitude $\eta(\br,t)$ and
the phase $\varphi(\br,t)$ are as follows
 \bee
   &&\eta(\br, t) =c_x(t)x+c_y(t)(y^2-z^2),
   \\
   &&\varphi(\br, t) =\displaystyle \br\tilde\Omega(t) \br+a(t)yz,
    \label{varable2}
\ene
 where as before $c_{x,y}(t)$ and $a(t)$ are
 functions of time   such that $c_x(t)c_y(t)>0$, and
 we have introduced the diagonal time-dependent 3$\times$3 matrix
 $\tilde\Omega=$diag$(\Omega_x,\Omega_y/2,\Omega_z/2)$.
   Now, we have $\rho(t)=\sqrt{c_x(t)c_y(t)}$,
and the linear and nonlinear potentials given by
 \bee
 \nonumber
 &&  v(\br,t) = \br \tilde A(t) \br+b_y(t) yz-\frac{\mu}{2} c_x^2(t),
  \\
  \nonumber
 &&  g(\br,t) =c_x^2(t)+4c_y^2(t)(y^2+z^2),
\ene where $b_y(t)$ is given by Eq. (\ref{B}), and we have
introduced the diagonal time-dependent 3$\times$3 matrix $\tilde
A=$diag$(A_x, C_{y}- c_x^2/2, C_{y}- c_x^2/2)$ with $A_x$ and
$C_{\sigma}$ being defined by Eq. (\ref{A}) and
 \bee \label{C}
C_{\sigma}=\frac{2c_{\sigma}(t)\ddot{c}_{\sigma}(t)-3\dot{c}_{\sigma}^2(t)}{8c_{\sigma}^2(t)}
-2\mu c_{\sigma}^2(t), \ene respectively.

 Like in the previous
case we observe that temporal evolution of the curves leads to
change of the phase surface, whose position becomes $x$-dependent
(c.f. the Case II in Table~\ref{surface}).

\subsection{Hyperboloid depending on time}
Here we consider the generalization of the hyperbolic Case III
from  Table~\ref{surface} for which the amplitude $\eta(\br,t)$
and phase $\varphi(\br,t)$ are of the forms
\bee
 &&  \eta(\br, t)=\br \hat c(t) \br,
 \\
 &&  \varphi(\br, t) =\frac 12  \br\hat\Omega(t)\br+a(t)xyz, \ \
\label{varable3b} \ene where as before $c_{\sigma}(t)$ and
$a (t)$ are functions of time $t$ such that
$c_x(t)c_y(t)c_z(t)>0$, and the condition
\bee
 \label{condition3}
 \mbox{Tr}\, \hat c(t)=0,
\ene is required. Moreover we have introduced the diagonal
time-dependent 3$\times$3 matrix $\hat c=$diag$(c_x, c_y, c_z)$.
Now we have $\rho(t)= [c_x(t)c_y(t)c_z(t)]^{1/4}$, and the
nonlinearity $g(\br,t)$ and potential $v(\br,t)$ are given by \bee
 \label{g3b}
  g(\br,t)&=& 4\br \hat c(t) \br,
 \\
  v(\br, t)&=&\br\hat C(t)\br-\left[a(t)\mbox{Tr}\,\hat\Omega(t)+\dot{a}(t)\right]xyz
  \nonumber
 \\
   && \quad+\frac{1}{2}a^2(t)(y^2z^2+x^2z^2+x^2y^2).
  \ene

 \subsection{Third order surface depending on time}

 Finally we consider the time-dependent generalization of the third order surface in Case IV from
the Table~\ref{surface} for which the amplitude $\eta(\br,t)$ and
phase $\varphi(\br,t)$ are as follows
\bee
   && \eta(\br, t)=c(t)xyz,
  \\\
  &&\varphi(\br, t)= \br\hat a(t) \br,
  \label{varable4}
  \ene
  where as before $c(t)$ and $a_{\sigma}(t)$ are
 functions of time $t$ such that $c_x(t)>0$ and the condition
 \bee
 \label{condition4} \dot{c}(t)+2c(t)\mbox{Tr}\,\hat a(t)=0,
  \ene
is required.  Here we have introduced the diagonal time-dependent
3$\times$3 matrix $\hat a=$diag$(a_x, a_y, a_z)$. Now
$\rho(t)=\sqrt{c (t)}$, the nonlinearity $g(\br,t)$ and potential
$v(\br,t)$ are as follows
 \bee
 \label{g4}
 && g(\br,t)=c^2(t)(y^2z^2+x^2z^2+x^2y^2), \\
 && v(\br,t)=  \br\hat D(t)\br -\frac{\mu}{2}
 c^2(t)(y^2z^2+x^2z^2+x^2y^2), \qquad
  \ene
where $\hat D=$diag$(D_x,D_y,D_z))$  with $D_{\sigma}=-\dot{a}_{\sigma}(t)-2a_{\sigma}^2(t)$.
Thus the introduced temporal dependence does not increase the orders of the potentials: both the linear
and nonlinear potentials are quartic.

\section{Generalized similarity reductions and solutions}

\subsection{The extension of the reduction equation}

The approach developed in the previous sections allows for further
generalizations. Indeed, it was based on the reduction of 3D models
to  1D NLS equations which admit  exact solutions. The latter however need not
 necessarily be equations with constant coefficients. They  can have
either linear and/or nonlinear inhomogeneous coefficients, however
still admitting exact solutions. Such situations are well known, for
example for the case of periodic coefficients~\cite{nontri,BK},
which can be constructed using the ``inverse engineering"
described in~\cite{BK}, and for localized and more sophisticated
spatial dependencies~\cite{Juan1}. Respectively, one can consider
  reductions of an  3D model to an {\em inhomogeneous}
1D equation with known solutions.

This leads us to the goal of this subsection: we intend to reduce Eq.
(\ref{5NLS}) to the stationary p-q NLS equation with the
$\eta$-modulated potentials $\mathcal{V}(\eta)$ and
$\mathcal{G}_{p,q}(\eta)$
 \begin{equation}
\mu\Phi=-\Phi_{\eta\eta}+\mathcal{V}(\eta)\Phi+\mathcal{G}_{p}(\eta)|\Phi|^{p-1}\Phi
+\mathcal{G}_{q}(\eta)|\Phi|^{q-1}\Phi, \label{ODEV}
\end{equation}
where $\Phi\equiv \Phi(\eta)$ is a function of the only
variable $\eta\equiv \eta(\br,t)$ whose relation to the original
variables $(\br,t)$ is to be determined, and $\mu$ is the eigenvalue
of the nonlinear equation.

We still consider the general similarity transformation
(\ref{Tran}). Insertion of Eq. (\ref{Tran}) into Eq. (\ref{5NLS}) and
requirement that $\Phi(\eta)$ satisfies Eq. (\ref{ODEV}) yield a set
of nonlinear partial differential equations, which for the
amplitude and phase surfaces coincide with previously obtained
ones (\ref{sys1})-(\ref{sys3}), and for the linear and nonlinear
potentials acquire the generalized form
\bee \label{Gpotential}
 \begin{array}{l}
  g_j(\br,t)=\displaystyle \frac{1}{2}\rho^{1-j}\mathcal{G}_j(\eta)|\nabla\eta|^2 \quad
  (j=p,\,q),  \vspace{0.08in}\cr
  v(\br,t)=\displaystyle\frac{1}{2}\Big[(V(\eta)-\mu)|\nabla\eta|^2-|\nabla\varphi|^2
   +\rho^{-1}\nabla^2\rho\Big]-\varphi_t,
  \end{array}  \ene

Now we can obtain the {\em new analytical solutions} of
Eq.(\ref{5NLS}) from those of Eq. (\ref{ODEV}) in terms of
similarity transformation (\ref{Tran}) and the corresponding
$\eta(\br, t)$ and $\varphi(\br, t)$ given in Sec. III and IV.

Passing to examples we restrict the consideration to the cubic case $\mathcal{G}_q\equiv 0$ and $p=3$ and
make two observations. First, being interested in flows, i.e. in solutions with varying phases and  choosing a
 periodic linear potential $\mathcal{V}(\eta)$ in a form of the elliptic function
 $
 \mathcal{V}(\eta)=-\mathcal{V}_0\ {\rm sn}^2(\omega \eta, k),
 $
where $\mathcal{V}_0$ and $\omega$ are constants, and $k\in [0,\
1]$ is the elliptic modulus,  one can construct new 3D solutions using the respective 1D problems intensively
studied in the literature (see e.g.~\cite{nontri,BK}).

Second, a new set of the solutions can be generated by the choice of the linear potential in the form
 \bee
  \mathcal{V}(\eta)=\omega^4\eta^2-\alpha^2\mathcal{G}_3(\eta)H_n^2(\omega\eta)e^{-\omega^2\eta^2},
   \ene
where $\alpha, \omega$ are constants, $\mathcal{G}_3(\eta)$ is an
arbitrary function of $\eta$, and $H_n(\omega\eta)$ is a Hermite
polynomial~\cite{Hermite}, then one obtaines the Hermite-Gaussian
solution of the nonlinear cubic Eq. (\ref{ODEV})
 \bee
  \label{SHG}
 \Phi(\eta)=\alpha H_n(\omega\eta)e^{-\omega^2\eta^2/2}, \quad \mu=\omega^2(2n+1).
\ene

The described solution allows for direct generalization to the
p-NLS case of arbitrary even $p\geq 4$ and $G_{q}=0$ for which
Eq.~(\ref{ODEV}) becomes
 \bee
    \mu\Phi=-\Phi_{\eta\eta}+\mathcal{V}(\eta)\Phi+\mathcal{G}_p(\eta)|\Phi|^{p-1}\Phi
  \ene
  whose Hermite-Gaussian solution is of the form (\ref{SHG})
  with $\mu=\omega^2(2n+1)$ and the chosen linear potential reads
$\mathcal{V}(\eta)=\omega^4\eta^2-\alpha^{p-1}\mathcal{G}_p(\eta)
 H_n^{p-1}(\omega\eta)e^{(1-p)\omega^2\eta^2/2}$.

\subsection{Generalized stationary reductions}

The assumption that $\Phi$ is given by the expression
(\ref{implicit}), does not necessarily requires that $\rho(\br)$
is a constant (as this has been assumed in
Sec.~\ref{sec:stationary}). Including the coordinate dependence in
the definition of $\rho$ represents another way of generalizing
the results obtained above. The respective results are obtained
directly from the equations (\ref{sys1})-(\ref{sys3}), giving that
now $\rho(\br)=\sqrt{1/f^\prime(\eta)}$ where $f(\eta)$ is an arbitrary
function having positive derivative, $f^\prime\equiv df/d\eta>0$ and $\eta$ is
given by one of the cases listed in the Table~\ref{surface}. Now  $\Phi$ is obtained
from the expression (\ref{implicit}), where $\eta$ is substituted
by $\tilde{\eta}=f(\eta)$.
Moreover, the corresponding phase $\varphi$ is as in the
stationary case (see the Table~\ref{surface}). This leads to the
obvious modifications of the linear and nonlinear potentials
directly following from Eqs. (\ref{sys4}) and (\ref{sys5}) (or Eq.
(\ref{Gpotential})) .

Inversely, since the equations (\ref{sys1})-(\ref{sys3}) for the
amplitude and phase surfaces are symmetric for
$\rho_t=\eta_t=\varphi_t=0$, for a given
$\rho(\br)=\sqrt{1/f^\prime(\varphi)}$, one can hold the same
amplitude surfaces, as in the stationary case (see the
Table~\ref{surface}), with the phase being chosen as
$\tilde{\varphi}=f(\varphi)$. Subsequently, this also leads to the
obvious modifications of the linear and nonlinear potentials
directly following from Eqs. (\ref{sys4}) and (\ref{sys5}) (or Eq.
(\ref{Gpotential})).

\subsection{Generalized time-dependent reductions}

In addition, if we consider the general case $\rho(\br, t)$ in Eq.
(\ref{Tran}) depending on both time $t$ and space $\br$, then
based on Eqs. (\ref{sys1})-(\ref{sys3}) we can obtain the general
amplitude $\eta(\br, t)$, the phase $\varphi(\br, t)$ and
$\rho(\br, t)$ listed in Table~\ref{surfaceg}, for which the
corresponding general linear and nonlinear potentials, i.e. $g_j(\br,
t)$ and $v(\br, t)$,
 can be obtained from Eq. (\ref{Gpotential}).
 Notice that the obtained $\eta(\br,t), \rho(\br, t), \ v(\br,t)$ and
$g_j(\br,t)$ all contain new arbitrary function $\Gamma(\zeta(\br,
t))$, but the corresponding phases $\varphi(\br,t)$ have no change
which are the same as ones in Sec. IV. Therefore the similarity
transformation (\ref{Tran}) containing the arbitrary function
$\Gamma(\zeta(\br, t))$ and solutions of Eq. (\ref{ODEV}) will
lead to the {\em abundant new solution profiles} of Eq.
(\ref{5NLS}).

\begin{table*}
\caption{Admissible $(\br, t)$-modulated amplitude and phase
surfaces and $\rho(\br, t)$ ($\Gamma(\zeta)$ being an arbitrary
differentiable function)}
 \begin{tabular}{llll} \hline\hline \\ [-2.0ex]
 Case \qquad \qquad & Amplitude surface \quad\quad\quad\quad\quad\quad\quad\qquad\qquad \qquad & Phase surface
 $\varphi(\br,t)$   is given by \quad\quad\quad   & Function $\rho(\br,t)$ \\ [1.0ex] \hline  \\ [-2.0ex]
 i & $\eta(\br,t)=\Gamma(\zeta), \ \ \zeta=\bc(t)\cdot\br$  & Eq.(\ref{varable1b}) with $\bc(t)\cdot \ba(t)=0 $ &
      $\sqrt{c_x(t)c_y(t)c_z(t)/\Gamma'(\zeta)}$ \\  [1.0ex]
 ii & $\eta(\br,t)=\Gamma(\zeta), \ \ \zeta=c_x(t)x +c_y(t)(y^2-z^2)$ & Eq.(\ref{varable2})  &
  $\sqrt{c_x(t)c_y(t)/\Gamma'(\zeta)}$  \\   [1.0ex]
 iii & $\eta(\br,t)=\Gamma(\zeta),\ \ \zeta=c_x(t) x^2+c_y(t) y^2+c_z(t)z^2 $ \quad\quad &
 Eq.(\ref{varable3b}) with Eq.(\ref{condition3})  &
 $\sqrt[4]{c_x(t)c_y(t)c_z(t)/\Gamma'^2(\zeta)}$  \\  [1.0ex]
 iv & $\eta(\br,t)=\Gamma(\zeta),\ \  \zeta=c_x(t)xyz$  &  Eq.(\ref{varable4}) with Eq.(\ref{condition4})  &
   $\sqrt{c_x(t)/\Gamma'(\zeta)}$  \\ [1.0ex] \hline\hline
\end{tabular}
\label{surfaceg}
\end{table*}


\section{Conclusions}

In the present work we have shown that   a large diversity of 3D NLS equations with coefficients
depending on time can be mapped by the proper similarity transformation into 1D models allowing for
 exact solution. In such  reductions the original coordinates in the 1+3-space are reduced to the set of
 one-parametric coordinate surfaces, whose parameter plays the role of the coordinate of the new 1D
 equation. When the obtained equation allows for exact solutions, the respective solutions of the original
 3D model can be constructed immediately using different types of the admissible coordinate surfaces.

We considered power surfaces, which give origin to parabolic and quartic linear and nonlinear potentials.
Such potentials are typical for the physical applications in the nonlinear optics and in the mean-field theory
of Bose-Einstein condensates, what determines the large range of the possible applications of the found solutions,
as well as of the method itself.

We also point out that not only the exact solutions itself represent the major interest. As the 3D equation is
reduce to the 1D model one can consider the existence of the solutions of the original model on the basis of
the known existence of the reduced equation. So, for example,  1D NLS equations with periodic
linear and nonlinear potentials ~\cite{BluKon} or with a parabolic linear~\cite{AlZez} potential  allow for
existence of various branches of the solutions (which however can be found only numerically). Each of the
 branches can be parameterized by the frequency (or energy, or chemical potential, depending on the applications),
 and this leads to a parametric set of the 3D NLS equations with inhomogeneous potentials allowing for either
 localized or periodic solutions. In addition, the reported method can be also extended to the 3D
(or $d$-dimensional) p-q NLS equation (or coupled p-q NLS
equations) with varying potentials, nonlinearities, dispersions
and gain/loss terms~\cite{PTK, Mal, RDP, Bel}.

We however, left open several relevant questions. Among them we mention the stability of the
obtained solutions which hardly can be implemented with the
framework of a general scheme, similar to one used to obtain the
solutions. We also did not discuss the relation between the self-similar flows obtained in the present paper and  collapsing solutions, for which the similarity transformation (usually referred to as lens transformation) appears to be a powerful tool~\cite{Berge}. The complexity of this last issue is determent by the presence of   inhomogeneous end even
time-dependent linear an nonlinear potentials, requiring further detail study.

\acknowledgments

Work of Z.Y. has been supported by the FCT SFRH/BPD/41367/2007 and
the NNFC (No.60821002).

\appendix

\section{Appendix}
\label{cub_quint}

For the sake of convenience here we present several exact
solutions of the cubic-quintic NLS equation (\ref{ODE}) with
$p=3,\,q=5$, which are listed in Table \ref{CQsolution}.
\begin{table*}
\caption{Solutions for the case cubic-quintic NLS equation
($k'^2=1-k^2$)}
 \begin{tabular}{llllll}
    \hline\hline \\ [-2.0ex]
    Case & $G_3$ & $G_5$ \quad & $\mu$  &  $\omega^2$ &  $\Phi(\eta)$  \\ [1.0ex] \hline \\ [-2.0ex]
 \quad 1 & $\ \ 1$ & $ -1$ & $3(4k^2+1)/(64k^2)$ & $3/(16k^2)$
         & $3/8\big[1+ {\rm cn}(\omega\eta, k)\big]^{1/2}$ \\ [2.0ex]
 \quad 2 & $\ \ 1$ & $ -1$ & $-2$ & $3/2$ &
           $ \displaystyle \sqrt{7} \cos(\omega\eta)\big[9-7\cos^2(\omega\eta)\big]^{-1/2}$  \\ [2.0ex]
 \quad 3 & $\ \ 1$ & $ -1$ & $\omega^2(2k^2-1)/2-5/4$
         & $ \big[2(1-2k^2)+\sqrt{4(1-2k^2)^2+45}\ \big]/6 $ &
           $\displaystyle\frac{\sqrt{10}k\ {\rm cn}(\omega\eta,k)}
             {\big[3(2\omega^2+6k^2-3)-10 k^2 \ {\rm cn}^2(\omega\eta, k)\big]^{1/2}}$ \\ [3ex]
 \quad 4 & $\ \ 1$ & $ -1$ & $0 $ & $ 1$ & $\sqrt{6}\ \big[4+3\eta^2\big]^{-1/2}$ \\ [1ex] \hline \\ [-1.0ex]
 \quad 5 & $-1$ & $ \ \ 1$ & $3(k^2-5)/64$ & $3/16$ &
           $3/8\big[1+k\ {\rm sn}(\omega\eta, k)\big]^{1/2}$ \\ [3.0ex]
 \quad 6 & $-1$ & $\ \ 1$ & $-\omega^2(k^2+1)/2+5/4$ \qquad &  $\big[-2(1+k^2)+\sqrt{4(1+k^2)^2+45k'^4}\ \big]/(6k'^4)$
  \quad &
           $\displaystyle \frac{\sqrt{10} k \ {\rm sn}(\omega\eta,k)}
            {\big[6\omega^2k'^4+9(k^2+1)-10 k^2 {\rm sn}^2(\omega\eta, k)\big]^{1/2}}$ \\ [3.0ex]
 \quad 7 & $-1$ & $\ \  1$ & $1/4$ & $ 1$ & $\eta \big[24+2\eta^2\big]^{-1/2}$ \\ [2.0ex]
 \hline\hline
\end{tabular}
\label{CQsolution}
\end{table*}



\end{document}